\documentstyle[12pt]{article}
\catcode`\@=11
\long\def\@makefntext#1{
\protect\noindent \hbox to 3.2pt {\hskip-.9pt
$^{{\ninerm\@thefnmark}}$\hfil}#1\hfill}		%CAN BE USED

 \def\@makefnmark{\hbox to 0pt{$^{\@thefnmark}$\hss}}  %ORIGINAL

\def\ps@myheadings{\let\@mkboth\@gobbletwo
\def\@oddhead{\hbox{}
\rightmark\hfil\ninerm\thepage}
\def\@oddfoot{}\def\@evenhead{\ninerm\thepage\hfil
\leftmark\hbox{}}\def\@evenfoot{}
\def\sectionmark##1{}\def\subsectionmark##1{}}

\newcounter{sectionc}\newcounter{subsectionc}\newcounter{subsubsectionc}
\renewcommand{\section}[1] {\vspace{0.6cm}\addtocounter{sectionc}{1}
\setcounter{subsectionc}{0}\setcounter{subsubsectionc}{0}\noindent
	{\bf\thesectionc. #1}\par\vspace{0.4cm}}
\renewcommand{\subsection}[1] {\vspace{0.6cm}\addtocounter{subsectionc}{1}
	\setcounter{subsubsectionc}{0}\noindent
	{\it\thesectionc.\thesubsectionc. #1}\par\vspace{0.4cm}}
\renewcommand{\subsubsection}[1] {\vspace{0.6cm}\addtocounter{subsubsectionc}
{1}
	\noindent {\rm\thesectionc.\thesubsectionc.\thesubsubsectionc.
	#1}\par\vspace{0.4cm}}

\newcounter{appendixc}
\newcounter{subappendixc}[appendixc]
\newcounter{subsubappendixc}[subappendixc]

\renewcommand{\appendix}[1] {\vspace{0.6cm}
        \refstepcounter{appendixc}
        \setcounter{figure}{0}
        \setcounter{table}{0}
        \setcounter{equation}{0}
        \renewcommand{\thefigure}{\Alph{appendixc}.\arabic{figure}}
        \renewcommand{\thetable}{\Alph{appendixc}.\arabic{table}}
        \renewcommand{\theappendixc}{\Alph{appendixc}}
        \renewcommand{\theequation}{\Alph{appendixc}.\arabic{equation}}
        \noindent{\bf Appendix \theappendixc #1}\par\vspace{0.4cm}}

\def\abstracts#1{{
	\centering{\begin{minipage}{30pc}\tenrm\baselineskip=12pt\noindent
	\centerline{\tenrm ABSTRACT}\vspace{0.3cm}
	\parindent=0pt #1
	\end{minipage}}\par}}

\renewenvironment{thebibliography}[1]
	{\begin{list}{\arabic{enumi}.}
	{\usecounter{enumi}\setlength{\parsep}{0pt}
\setlength{\leftmargin 1.25cm}{\rightmargin 0pt}
	 \setlength{\itemsep}{0pt} \settowidth
	{\labelwidth}{#1.}\sloppy}}{\end{list}}

\newcounter{itemlistc}
\newcounter{romanlistc}
\newcounter{alphlistc}
\newcounter{arabiclistc}

\newcommand{\fcaption}[1]{
        \refstepcounter{figure}
        \setbox\@tempboxa = \hbox{\tenrm Fig.~\thefigure. #1}
        \ifdim \wd\@tempboxa > 6in
           {\begin{center}
        \parbox{6in}{\tenrm\baselineskip=12pt Fig.~\thefigure. #1}
            \end{center}}
        \else
             {\begin{center}
             {\tenrm Fig.~\thefigure. #1}
              \end{center}}
        \fi}

\newcommand{\tcaption}[1]{
        \refstepcounter{table}
        \setbox\@tempboxa = \hbox{\tenrm Table~\thetable. #1}
        \ifdim \wd\@tempboxa > 6in
           {\begin{center}
        \parbox{6in}{\tenrm\baselineskip=12pt Table~\thetable. #1}
            \end{center}}
        \else
             {\begin{center}
             {\tenrm Table~\thetable. #1}
              \end{center}}
        \fi}

\def\@citex[#1]#2{\if@filesw\immediate\write\@auxout
	{\string\citation{#2}}\fi
\def\@citea{}\@cite{\@for\@citeb:=#2\do
	{\@citea\def\@citea{,}\@ifundefined
	{b@\@citeb}{{\bf ?}\@warning
	{Citation `\@citeb' on page \thepage \space undefined}}
	{\csname b@\@citeb\endcsname}}}{#1}}

\newif\if@cghi
\def\cite{\@cghitrue\@ifnextchar [{\@tempswatrue
	\@citex}{\@tempswafalse\@citex[]}}
\def\citelow{\@cghifalse\@ifnextchar [{\@tempswatrue
	\@citex}{\@tempswafalse\@citex[]}}
\def\@cite#1#2{{$\null^{#1}$\if@tempswa\typeout
	{IJCGA warning: optional citation argument
	ignored: `#2'} \fi}}

\def\fnt#1#2{\footnotetext{\kern-.3em
	{$^{\mbox{\sevenrm #1}}$}{#2}}}

 1
 1
 1

\font\tenbf=cmbx10
\font\tenrm=cmr10
\font\tenit=cmti10

\font\ninerm=cmr9

\evensidemargin
0.30truein
\begin{document}

\centerline{\tenbf HEAVY MESON DYNAMICS}
\centerline{\tenbf IN A QCD RELATIVISTIC POTENTIAL MODEL}
\vspace{0.8cm}
\centerline{\tenrm FULVIA DE FAZIO}
\baselineskip=13pt
\centerline{\tenit Dipartimento di Fisica dell'Universit\'a di Bari,}
\baselineskip=12pt
\centerline{\tenit Istituto Nazionale di Fisica Nucleare, Sezione di Bari,}
\centerline{\tenit Via Amendola 173, 70126, Bari, Italy}
\vspace{0.3cm}
\vspace{0.9cm}
\abstracts{We use a QCD relativistic potential model to compute
the strong coupling constant $g$ appearing in the effective Lagrangian
which describes the interaction of $0^-$ and $1^-$ $\bar q Q$
states  with soft pions in the limit $m_Q \to \infty$.
We compare our results with other approaches; in particular, in the non
relativistic limit, we are able to
reproduce the constituent quark model result: $g=1$, while the inclusion of
relativistic effects due to the light quark gives $g={1 \over 3}$, in
agreement with  QCD sum rules.
We also estimate heavy meson radiative decay rates, with results in agreement
with available experimental data.}

\vfil
%\vspace{0.8cm}
\rm\baselineskip=14pt
\section{The Strong Coupling Constant $g_{D^*D\pi}$}
The decay $D^{*+} \to D^0\pi^+$ is described in terms of a strong coupling
constant $g_{D^*D\pi}$ defined by:
$<D^0(k)\pi^+(q)|D^{*+}(p,\epsilon)>\;=g_{D^*D\pi} \; \epsilon^\mu q_\mu $.
CLEO  collaboration measurement \cite{CLEO}:
$BR(D^{*+} \to D^0\pi^+) \simeq 68.1 \pm 1.0 \pm 1.3 \%$ and the
upper bound\cite{ACCMOR} : $\Gamma(D^{*+})< 131 \; KeV$ provide
us with the constraint : $g_{D^*D \pi} < 20.6$.\par
The interest in the evaluation of this coupling constant is manifold. The form
factor $F_1(q^2)$ describing the semileptonic decay $B \to \pi \ell \nu$ is
believed to be dominated by the $B^*$ pole, so that its value at maximum
transferred momentum  is proportional to $g_{B^*B\pi}$, which is
 related to $g_{D^*D \pi}$ by \cite{pham}:
$g_{P^*P \pi}=2 \; m_P  \; g / f_\pi   \hskip 3 pt , $
where $P(P^*)$ is a $0^-$ ($1^-$) heavy meson of mass $m_P$,
and $g$ is independent of $m_P$.
Besides, $g$ appears in the effective Lagrangian  describing the
interaction between heavy mesons and light Nambu-Goldstone bosons
\cite{wise,burdman,yan}.\par
In non relativistic quark models \cite{yan,isgur}  $g \simeq 1$, while
recent QCD sum rules \cite{colangelo} and HQET \cite{casal,noi}
analyses give: $g \simeq 0.2 - 0.4$. We wish to show
that the inclusion of the relativistic effects in the bound state can lower the
value $g=1$, providing  an
explanation of the discrepancy between the different approaches.\par
We shall obtain $g$ in the framework of a QCD relativistic potential model
\cite{nostro}. In this model
the ${\bar q}Q$ heavy states $D$ and $D^*$  are
described in terms of the creation operators  of the constituent quarks and of
a meson wave function $\psi$, normalized according to:
${1 \over (2\pi)^3} \int d \vec{k} |\psi|^2=2\sqrt{m_D^2+\vec{p}^2}$, where
$\vec{p}$ is the meson momentum and $\vec{k}$ is the quark relative momentum.
$\psi$ satisfies a
Salpeter equation\footnote[1]{This equation arises from
the bound-state Bethe Salpeter equation by considering the instantaneous
approximation and restricting the Fock space to the ${\bar q}Q$ pairs
\cite{pietroni}.} which includes relativistic effects in the
quark kinematics. In this equation the interquark potential $V$ coincides,
in the meson rest
frame, with the Richardson potential,  with a  linear behaviour
for large distances, reproducing QCD confinament, and a coulombic shape
for small distances, with the further assumption that $V$ is constant near
the origin in order to avoid unphysical singularities.
\noindent The values of the quark masses, obtained by fits to meson masses,
are:
$m_u=m_d=38 \hskip 3 pt MeV$, $m_s=115 \hskip 3 pt  MeV$, $m_c=1452
\hskip 3 pt MeV$, $m_b=4890 \hskip 3 pt MeV$. \par
To evaluate $g_{D^*D\pi}$ let us consider the matrix element of the
axial current $A_\mu$ between the states $D^*$ and $D$:
\begin{eqnarray}
<D^0(k) | A_\mu | D^{*+}(p, \epsilon) > =-i \{&\epsilon_\mu&(m_{D^*}+m_D)
A_1(q^2)-{\epsilon \cdot q \over m_D + m_{D^*}}(p+k)_\mu A_2(q^2)  \nonumber \\
&-& {\epsilon \cdot q \over q^2} 2 m_{D^*} q_\mu [A_3(q^2)-A_0(q^2)] \} \hskip
 3 pt , \label{eq : 1} \end{eqnarray}
\noindent where $2m_{D^*} A_3=(m_D + m_{D^*})A_1+(m_{D^*}-m_D)A_2$ $\;\;$
($q=p-k$).\par
Taking the derivative of $A_\mu$, we can link the l.h.s. of Eq.~(\ref{eq : 1})
to the matrix element of the pseudoscalar current between the same states,
which is supposed to be dominated by the $\pi^+$ pole.
Performing the overlap of
the states $D$, $D^*$ and the axial current, we have in the chiral limit:
$g_{D^* D \pi} = {2 m_{D^*} \over f_\pi} A_0(0) $, and finally:
\begin{equation}
 g=A_0(0)=
{1 \over 4 m_D} \int_0^\infty dk |\tilde{u}(k)|^2 {E_q + m_q
\over E_q} \left[ 1-{k^2 \over 3 (E_q+m_q)^2} \right] \hskip 3 pt ,
\label{eq : 2} \end{equation}
\noindent where $E_q=\sqrt{k^2+m_q^2}$ and
$\tilde{u}(k)={k \; \psi(k) \over \sqrt{2} \pi }$. The wave functions
$\tilde{u}(k)$ come from a numerical solution of the Salpeter equation.
We obtain \cite{nostro}:
$A_0(0)=0.4  \hskip 0.5 cm (D  \;\;\; case)$;
$A_0(0)=0.39 \hskip 0.5 cm (B  \;\;\; case)$,
giving so: $g_{D^*D \pi}=12.3$ and $g_{B^*B \pi}=31.7$. We can
observe that we have only $2\%$ deviation from the scaling result $g_{D^*D \pi}
/g_{B^*B \pi}=m_D/m_B$.\par
It is interesting to notice that, in the non relativistic
limit, i.e. $E_q \simeq m_q \gg k$, we obtain:
$g={1 \over 2 m_D} \int_0^\infty dk |\tilde{u}(k)|^2=1 $,
reproducing the constituent quark model result. On the other hand,
in the limit $m_q \to 0$, $m_Q \to \infty$, the result is: $g=1/3$,
in agreement with the QCD sum rules determination.

\section{Radiative Heavy Meson Decays}
The evaluation of radiative decay rates involves the knowledge of the matrix
element of the electromagnetic current $J_\mu^{e.m.}$ between the states $D^*$
and $D$:

\begin{equation}
 <D^+(k)| J^{e.m.}_\mu|D^{*+}(p,\epsilon)>=\left( {e_Q \over \Lambda_Q} +
{e_q \over \Lambda_q}\right) \;
\epsilon_{\mu \nu \alpha \beta} \; \epsilon^\nu k^\alpha p^\beta
\hskip 3 pt , \label{eq : 3} \end{equation}

\noindent where  $e_Q$ $
(e_q)$ is the heavy (light) quark electric charge. In the framework of the
relativistic QCD model we are able to evaluate the "effective" masses
$\Lambda_q$, $\Lambda_Q$ by computing the
overlap of meson states and the electromagnetic current. We find
$\Lambda_c \simeq m_c=1.57 \hskip 3 pt GeV$, $\Lambda_b \simeq m_b =4.95
\hskip 3 pt GeV$ and $\Lambda_q=0.48 \hskip 3 pt GeV \gg m_q$ .
The B.R. are reported in Table I  \cite{nostro}
together with the available experimental data.

\begin{table}
\begin{center}
{\tcaption  {Heavy meson radiative decay rates}}
\begin{tabular}{l c c }
 \hline \hline
Decay rate/ BR & theory & experiment \\ \hline
$\Gamma(D^{*+})$ & $46.21  \hskip 3 pt KeV$ & $<$ 131 \hskip 5 pt $KeV $
 \\ \hline
$ BR(D^{*+} \to D^+ \pi^0)$ & $31.3 \%$ & $30.8 \pm 0.4 \pm 0.8 \%$ \\
\hline
$BR(D^{*+} \to D^0 \pi^+)$ & $67.7 \%$ & $68.1 \pm 1.0 \pm 1.3 \%$ \\
\hline
$BR(D^{*+} \to D^+ \gamma)$ & $1.0 \%$ & $1.1 \pm 1.4 \pm 1.6 \%$ \\
\hline
$\Gamma(D^{*0})$ & $41.6 \hskip 3 pt KeV$ &  \\ \hline
$ BR(D^{*0} \to D^0 \pi^0)$ & $50.0 \%$ & $63.6 \pm 2.3 \pm 3.3 \%$ \\
\hline
$BR(D^{*0} \to D^0 \gamma)$ & $50.0  \%$ & $36.4 \pm 2.3 \pm 3.3 \%$ \\
\hline
$\Gamma(D^*_s)=
\Gamma(D^*_s \to D_s \gamma)$ & $ 0.382  \hskip 3  pt KeV$ &  \\ \hline
$\Gamma(B^{*-})=\Gamma(B^{*-} \to B^- \gamma)$ & $0.243
\hskip 3  pt KeV$ &  \\ \hline
$\Gamma(B^{*0})=\Gamma(B^{*0} \to B^0 \gamma)$ & $9.2 \hskip 3 pt 10^{-2}
\hskip 3  pt KeV$ &  \\ \hline
$\Gamma(B^*_s)=\Gamma(B^*_s \to B_s \gamma)$ & $8.0 \hskip 3pt 10^{-2}
\hskip 3  pt KeV$ &  \\ \hline  \hline

\end{tabular}
\end{center}
\end{table}

We conclude that
the inclusion of relativistic effects in a QCD  potential model
has allowed us to explain the discrepancy among different models in the
evaluation of the heavy meson coupling constant with soft pions. Moreover,
the same model, applied to
heavy meson radiative transitions, gives results in agreement with
experimental data.\\

I thank P. Colangelo and G. Nardulli for their precious collaboration.

\end{document}